\documentclass[a4paper,fleqn,usenatbib]{mnras}

\usepackage[T1]{fontenc}
\usepackage{ae,aecompl}

\usepackage{graphicx}	
\usepackage{amsmath}	
\usepackage{amssymb}	
\usepackage{float}
\usepackage{natbib}
\usepackage{morefloats}
\usepackage{dcolumn}
\usepackage{amsmath}
\usepackage{latexsym}
\usepackage{longtable}
\usepackage{lipsum}
\usepackage{gensymb} 
\usepackage{adjustbox}
\usepackage{lscape}

\title[SN 2020rea]{Optical studies of a bright Type Iax supernova SN~2020rea}
\author[Mridweeka Singh et al.]{Mridweeka Singh$^{1}$\thanks{E-mail: mridweeka.singh@iiap.res.in, yashasvi04@gmail.com},
Kuntal Misra$^{2}$,
Devendra K. Sahu$^{1}$,
Bhavya Ailawadhi$^{2,3}$,
\newauthor
Anirban Dutta$^{1,4}$,
D. Andrew Howell$^{5,6}$,
G. C. Anupama$^{1}$,
K. Azalee Bostroem$^{7}$,
\newauthor
Jamison Burke$^{5,6}$,
Raya Dastidar$^{8,9}$,
Anjasha Gangopadhyay$^{10}$,
Daichi Hiramatsu$^{5,6,11,12}$,
\newauthor
Hyobin Im$^{13,14}$,
Curtis McCully$^{5,6}$,
Craig Pellegrino$^{5,6}$,
Shubham Srivastav$^{15}$,
\newauthor
Rishabh Singh Teja$^{1,4}$
\\
$^{1}$Indian Institute of Astrophysics, II Block, Koramangala, Bengaluru 560 034, India \\
$^{2}$Aryabhatta Research Institute of observational sciencES, Manora Peak, Nainital 263 001, India\\
$^{3}$Deen Dayal Upadhyaya Gorakhpur University, Gorakhpur 273009, India\\
$^{4}$Pondicherry University, R.V. Nagar, Kalapet, 605014, Puducherry, India\\
$^{5}$Las Cumbres Observatory, 6740 Cortona Drive, Suite 102, Goleta, CA 93117-5575, USA\\
$^{6}$Department of Physics, University of California, Santa Barbara, CA 93106-9530, USA\\
$^{7}$DiRAC Institute, Department of Astronomy, University of Washington, Box 351580, U.W., Seattle, WA 98195, USA\\
$^{8}$Millennium Institute of Astrophysics (MAS), Nuncio Monsenor Sòtero Sanz 100, Providencia, Santiago, Chile\\
$^{9}$Departamento de Ciencias Fisicas, Universidad Andres Bello, Fernandez Concha 700, Las Condes, Santiago, Chile\\
$^{10}$Hiroshima Astrophysical Science Center, Hiroshima University, Higashi-Hiroshima, Japan\\
$^{11}$ Center for Astrophysics \textbar{} Harvard \& Smithsonian, 60 Garden Street, Cambridge, MA 02138-1516, USA\\
$^{12}$The NSF AI Institute for Artificial Intelligence and Fundamental Interactions\\
$^{13}$Korea Astronomy and Space Science Institute, 776 Daedeokdae-ro, Yuseong-gu, Daejeon 34055, Republic of Korea\\
$^{14}$Korea University of Science and Technology (UST), 217 Gajeong-ro, Yuseong-gu, Daejeon 34113, Republic of Korea\\
$^{15}$Astrophysics Research Centre, School of Mathematics and Physics, Queen’s University Belfast, Belfast BT7 1NN, UK
}

\date{Accepted XXX. Received YYY; in original form ZZZ}

\pubyear{2022}

\begin{document}
\label{firstpage}
\pagerange{\pageref{firstpage}--\pageref{lastpage}}
\maketitle


\begin{abstract}
We present optical photometric and spectroscopic analysis of a Type Iax supernova (SN) 2020rea situated at the brighter luminosity end of Type Iax supernovae (SNe). The light curve decline rate of SN~2020rea is $\Delta$m$_{15}$(g) = 1.31$\pm$0.08 mag which is similar to SNe 2012Z and 2005hk. Modelling the pseudo bolometric light curve with a radiation diffusion model yields a mass of $^{56}$Ni of 0.13$\pm$0.01 M$_{\odot}$ and an ejecta mass of  0.77$^{+0.11}_{-0.21}$ M$_{\odot}$. Spectral features of SN~2020rea during the photospheric phase show good resemblance with SN 2012Z. TARDIS modelling of the early spectra of SN~2020rea reveals a dominance of Iron Group Elements (IGEs). The photospheric velocity of the Si {\sc II} line around maximum for SN~2020rea is $\sim$ 6500 km s$^{-1}$ which is less than the measured velocity of the Fe {\sc II} line and indicates significant mixing. The observed physical properties of SN~2020rea match with the predictions of pure deflagration model of a Chandrasekhar mass C-O white dwarf. The metallicity of the host galaxy around the SN region is 12+log(O/H) = 8.56$\pm$0.18 dex which is similar to that of SN 2012Z.  

\end{abstract}

\begin{keywords}
supernovae: general -- supernovae: individual: SN 2020rea --  galaxies: individual: UGC 10655   -- techniques: photometric -- techniques: spectroscopic 
\end{keywords}



\section{Introduction}
\label{Introduction}

Type Iax supernovae (SNe) are low luminosity and less energetic cousins of Type Ia SNe \citep{2003PASP..115..453L,2013ApJ...767...57F}. Type Iax SNe are known to have a wide range of luminosities (M$_{r}$ = $-$12.7 mag, \citealt{Karambelkar_2021} to M$_{V}$ = $-$18.4 mag, \citealt{2011ApJ...731L..11N}). There are bright members such as SNe 2011ay \citep{2015MNRAS.453.2103S,2017MNRAS.471.4865B}, 2012Z \citep{2015A&A...573A...2S} and faint members like SNe~2008ha \citep{2009AJ....138..376F, 2009Natur.459..674V}, 2010ae \citep{2014A&A...561A.146S}, 2019gsc \citep{2020ApJ...892L..24S, 2020MNRAS.496.1132T} and 2021fcg \citep{Karambelkar_2021}. However, dominance of relatively faint Type Iax SNe can be seen over bright ones  \citep{2011MNRAS.412.1441L, 2017ApJ...837..121G}. Though the sample size of Type Iax SNe is increasing with new discoveries by ongoing transient surveys, the progenitor and explosion mechanism of these peculiar objects are still debated. In order to understand them in a better way, detailed study of individual candidates is important. 

The pre-maximum spectra of Type Iax SNe are dominated by Intermediate Mass Elements (IMEs), Iron Group Elements (IGEs), along with C and O features. The pre-maximum spectral features are similar to SN 1991T-like Type Ia SNe \citep{2013ApJ...767...57F, 2014ApJ...786..134M} with weak Si {\sc II}, S {\sc II}, Ca {\sc II} lines and strong high excitation features such as Fe {\sc III}.  Measured expansion velocities of Type Iax SNe close to maximum lie between 2000 km s$^{-1}$ to 8000 km s$^{-1}$ \citep{2009AJ....138..376F,2014A&A...561A.146S} which is significantly less than the expansion velocities associated with Type Ia SNe ($\sim$ 11000 km s$^{-1}$,  \citealt{Wang_2009,2013ApJ...767...57F}). Type Iax SNe show different spectroscopic behaviour, especially at nebular phase with presence of permitted Fe {\sc II} lines \citep{2008ApJ...680..580S,2017hsn..book..375J}.  

The progenitor system of these explosions are not yet fully understood. Deep pre-explosion images are available for a few Type Iax SNe.  In the case of SN 2012Z, the analysis of the pre-explosion image led \cite{2014Natur.512...54M} to suggest  that the most favoured progenitor of this class could be a white dwarf in a binary system with Helium star as a companion. Nevertheless, the possibility of a single star as the progenitor was not completely ruled out in their work. Based on the pre-explosion images of SN 2014dt, \cite{2015ApJ...798L..37F} suggested that a C-O white dwarf in association with a Helium star can be a plausible progenitor system. Moreover, possible detection of Helium features in SNe 2004cs and 2007J were presented by \cite{2013ApJ...767...57F}. Detailed spectroscopic studies for a sample of Type Iax SNe, however, resulted in null detection of Helium feature \citep{2015ApJ...799...52W,2019MNRAS.487.2538J,2019A&A...622A.102M}. Hence, binary system with a Helium star companion of the progenitor white dwarf is still debated.
 
 The low luminosity and less energetic nature of Type Iax SNe suggest an incomplete disruption of the white dwarf which could lead to  a bound remnant. 
 The presence of P-Cygni lines and forbidden lines in the late phase spectra has been attributed to the centrally located optically thick high density region and optically thin SN ejecta, respectively \citep{2006AJ....132..189J,2008ApJ...680..580S} suggesting two component structure of the ejecta. \cite{2014ApJ...792...29F} presented late time observations of SN 2008ha and discussed about the possibilities of the remnant detection. The observed IR excess seen in the late time light curves of SN 2014dt  \citep{2016ApJ...816L..13F} was explained as arising from a bound remnant with an extended optically thick super-Eddington wind. Based on the late phase spectroscopic study for a larger sample, \cite{2016MNRAS.461..433F} have also proposed a two component model for the ejecta of SNe Iax. The possibility of the presence of a bound remnant in these explosions has also been discussed in  \cite{ 2014ApJ...786..134M,Shen_2017,Vennes680,2018PASJ...70..111K,Shen_2018,2019MNRAS.489.1489R,10.1093/pasj/psab075} and \cite{2022ApJ...925..138M}.

 \cite{2012ApJ...761L..23J}, \cite{2013MNRAS.429.2287K} and \cite{2014MNRAS.438.1762F} gave different deflagration models of C-O white dwarfs and could reproduce most of the observed features seen in relatively bright Type Iax SNe. A disk detonation associated with the merger of a white dwarf with a neutron star or black hole \citep{Fern_ndez_2013} can account for some properties seen in Type Iax SNe. On the other hand, to explain the observed properties of faint Type Iax SNe, several channels e.g. mergers involving C-O and O-Ne white dwarfs \citep{2018ApJ...869..140K}, partial deflagration associated with hybrid C-O-Ne white dwarf \citep{2015MNRAS.447.2696D,2015MNRAS.450.3045K,2016A&A...589A..38B}, deflagrations of C-O white dwarfs \citep{2022A&A...658A.179L}, core collapse scenario \citep{2010ApJ...719.1445M}, O-Ne white dwarf and neutron star/black hole mergers \citep{2022MNRAS.510.3758B}, and electron capture SN scenario \citep{Pumo_2009} have been proposed.

In this paper we present photometric and spectroscopic analysis of a bright Type Iax SN~2020rea. Section \ref{Discovery, observation and data reduction} mentions the discovery, follow-up and procedures used to reduce the data of SN~2020rea. A short description on the adopted distance and extinction is presented in Section \ref{distance_extinction}. In Section \ref{analaysis_light_curve}, the photometric properties of SN~2020rea are discussed. The bolometric light curve, its fitting with analytical models to infer the explosion parameters, and the comparison with deflagration models are presented in Section \ref{bolometric_light_curve}. 
Section \ref{spectral_properties} provides spectral studies of SN~2020rea and its host galaxy. A comparison of the observed features of SN~2020rea with the proposed explosion scenario for SNe Type Iax is made in Section \ref{explosion_sscenario}. Finally, a summary of this study is presented at the end of the paper in Section \ref{summary}.

\begin{table}
\caption{SN~2020rea and its host galaxy UGC 10655 }
\centering
\smallskip
\begin{tabular}{l l}
\hline \hline
\\
Host galaxy$^\star$ & UGC 10655    \\
Galaxy Morphology & Sbc \\  
Redshift & 0.02869$\pm$0.00015$^\dagger$ \\ 
Helio. Radial Velocity &  8600.15$\pm$44.07 km/sec \\
\\
\hline
\\
R.A.(J2000.0) & 16$^h$59$^m$37.82$^s$ \\
Dec.(J2000.0) & 56$^o$04$'$08.48${''}$ \\
Galactic extinction E(B-V) & 0.02 mag \\
Host extinction E(B-V) & 0.08 mag$^\ddagger$ \\
SN type & Iax\\
Offset from nucleus & 1$^{''}$.08 S 14$^{''}$.59 E \\
Date of Discovery & 2020-08-11 \\
\hline 
\end{tabular}
\newline \newline
$^\star$ The host galaxy parameters are taken from NED
\newline
$^\dagger$ \cite{1999PASP..111..438F} 
$^\ddagger$ See Section \ref{distance_extinction}
\label{tab:SN2020rea_detail}     
\end{table}

\section{Discovery, observation and data reduction}
\label{Discovery, observation and data reduction}

\begin{figure}
	\begin{center}
		\includegraphics[width=\columnwidth]{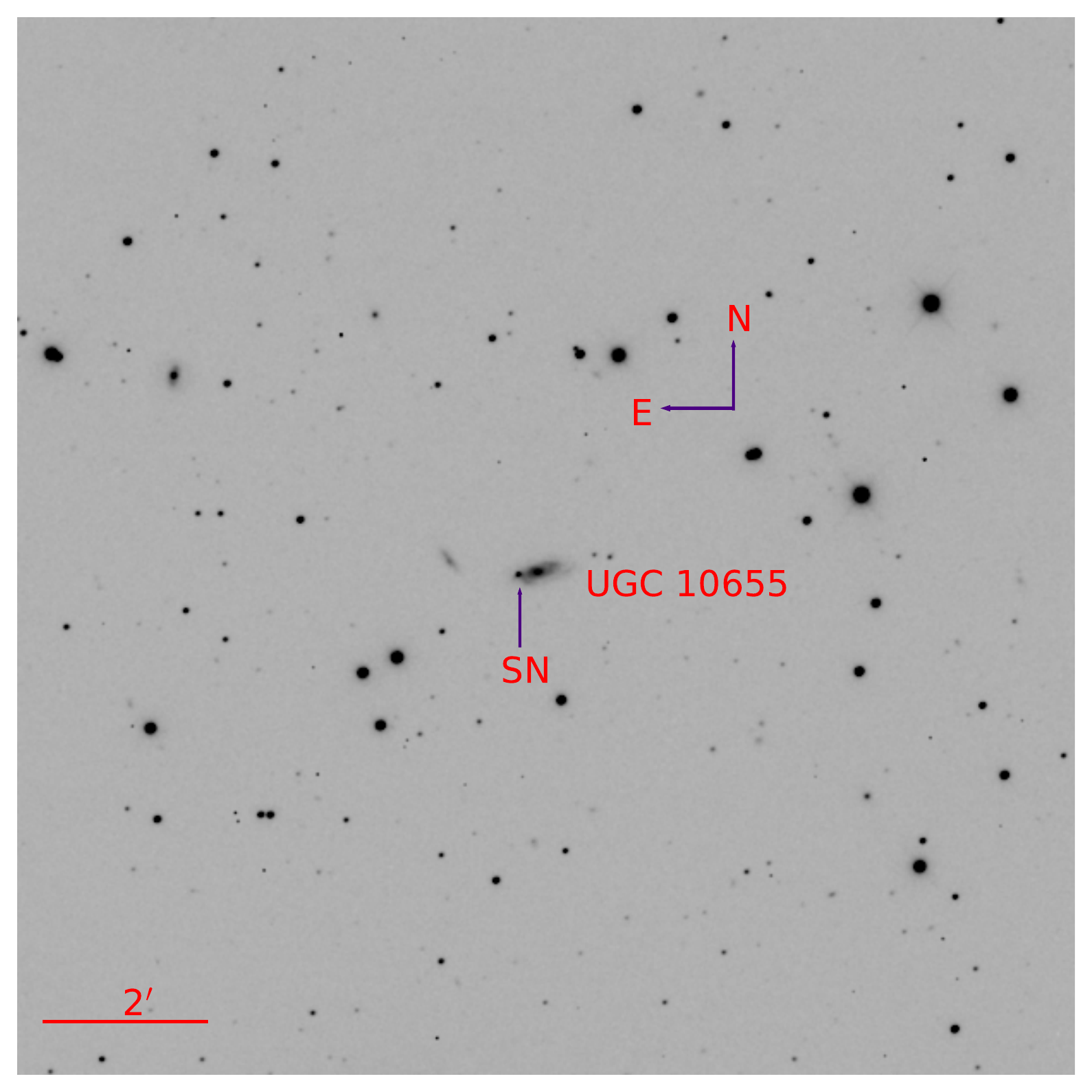}
	\end{center}
	\caption{Location of SN~2020rea in UGC 10655. This image is acquired on August 22, 2020 in {\it V}-band with 1m LCO telescope.}
	\label{fig:ds9_2020rea}
\end{figure}

\begin{table*}
\caption{Optical photometric observations of SN 2020rea}
\centering
\smallskip
\scriptsize
\begin{tabular}{c c c c c c c c  }
\hline \hline

Date          &   JD$^\dagger$  & Phase$^\ddagger$     &   B       	 &         V           	  &       g               &        r   	            &   i                   \\
              &                 &                     & (mag)     	 &        (mag)           &      (mag)            &       (mag) 	    &    (mag)            \\
\hline  
         
2020-08-17     &    78.75   & -5.98  &   --     --	     &    18.03 $\pm$ 0.05 &	 17.79 $\pm$ 0.04 &	 17.95 $\pm$ 0.04    &  18.23 $\pm$  0.03    \\
2020-08-23     &    84.73   & 0.00   &  17.49 $\pm$ 0.03 &    17.40 $\pm$ 0.03 &	 17.34 $\pm$ 0.03 &	 17.39 $\pm$ 0.02    &  17.65 $\pm$  0.03    \\
2020-09-04     &    96.71   & 11.97  &  18.55 $\pm$ 0.07 &    17.67 $\pm$ 0.03 &	 18.20 $\pm$ 0.04 &	 17.34 $\pm$ 0.02    &  17.51 $\pm$  0.02    \\
2020-09-08     &    100.62  & 15.89  &  19.27 $\pm$ 0.05 &    18.00 $\pm$ 0.04 &	 18.90 $\pm$ 0.06 &	 17.79 $\pm$ 0.03    &  17.60 $\pm$  0.01    \\
2020-09-12     &    104.63  & 19.99  &  19.77 $\pm$ 0.17 &    18.28 $\pm$ 0.05 &	 19.30 $\pm$ 0.10 &	 17.81 $\pm$ 0.04    &  17.72 $\pm$  0.02    \\
2020-09-18     &    110.68  & 25.94  &  20.29 $\pm$ 0.32 &    18.59 $\pm$ 0.07 &	 19.66 $\pm$ 0.10 &	 18.10 $\pm$ 0.05    &  18.01 $\pm$  0.02    \\
2020-09-22     &    114.67  & 29.93  &  20.24 $\pm$ 0.10 &    18.76 $\pm$ 0.09 &	 19.65 $\pm$ 0.11 &	 18.31 $\pm$ 0.06    &  18.15 $\pm$  0.03    \\
2020-09-26     &    118.63  & 33.89  &  20.88 $\pm$ 0.39 &    19.03 $\pm$ 0.09 &	 19.70 $\pm$ 0.14 &	 18.54 $\pm$ 0.05    &  18.37 $\pm$  0.04    \\
2020-09-30     &    122.63  & 37.90  &  20.76 $\pm$ 0.40 &    19.16 $\pm$ 0.09 &	 19.91 $\pm$ 0.13 &	 18.66 $\pm$ 0.08    &  18.50 $\pm$  0.05    \\
2020-10-07     &    129.61  & 44.88  &  20.90 $\pm$ 0.34 &    19.21 $\pm$ 0.12 &	 19.94 $\pm$ 0.16 &	 18.86 $\pm$ 0.13    &  18.87 $\pm$  0.05    \\
2020-10-14     &    136.55  & 51.81  &  20.90 $\pm$ 0.39 &    19.44 $\pm$ 0.10 &	 20.29 $\pm$ 0.22 &	 19.07 $\pm$ 0.13    &  18.91 $\pm$  0.05    \\
2020-10-15     &    137.55  & 52.81  &  20.49 $\pm$ 0.32 &    19.49 $\pm$ 0.14 &	 20.11 $\pm$ 0.16 &	 18.98 $\pm$ 0.10    &  18.98 $\pm$  0.06    \\
2020-10-20     &    142.55  & 57.81  &  20.73 $\pm$ 0.44 &    19.34 $\pm$ 0.12 &	 20.34 $\pm$ 0.21 &	 19.17 $\pm$ 0.13    &  19.11 $\pm$  0.08    \\
2020-11-02     &    155.54  & 70.81  &  --      --       &    20.02 $\pm$ 0.43 &	 --      --       &	 --      --    	     &  --      --        \\
2021-01-05     &    220.03  & 135.29 &  --      --       &    20.30 $\pm$ 0.26 &     20.49 $\pm$ 0.19 &  20.34 $\pm$ 0.26    &  20.11 $\pm$  0.14    \\
2021-01-09     &    224.01  & 139.27 &  20.61 $\pm$0.24  &    20.18 $\pm$ 0.17 &         --      --   &      --      --      &  --      --        \\ 

\hline \hline 
\end{tabular}  
\newline
\flushleft
\vspace*{-0.1cm}
\hspace*{1.5cm}
$^\dagger$2459000+
\newline
\flushleft
\vspace*{-0.5cm}
\hspace*{1.5cm}
$^\ddagger$with respect to  g$_{max}$= 2459084.74                    
\label{tab:photometric_observational_log_2020rea}                                                        
\end{table*}

\begin{table}
\caption{Log of spectroscopic observations}
\centering
\smallskip
\begin{tabular}{c c c }
\hline \hline
Date          & Phase$^\dagger$        & Telescope/Instrument       \\
              &(Days)                  &                          \\
\hline
2020-08-16      & -7.0      &    FTN/FLOYDS \\
2020-08-19      &  -4.0     &    FTN/FLOYDS  \\
2020-08-22      & -0.9  	&    FTN/FLOYDS  \\
2020-08-23      & 0.0      &    FTN/FLOYDS \\
2020-09-02      &  9.9      &    FTN/FLOYDS  \\
2020-09-13      & 20.9 	    &    FTN/FLOYDS  \\
\hline                          
\end{tabular}
\newline
$^\dagger$ Phase is calculated with respect to  g$_{max}$= 2459084.74
\label{tab:spectroscopic_observations_20rea}      
\end{table}

SN~2020rea was spotted by Supernova and Gravitational Lenses Follow up (SGLF) team in the Zwicky Transient Facility (ZTF) data \citep{2020TNSTR2463....1P} on August 11, 2020 (JD=2459072.702) in the host galaxy UGC 10655 at a redshift of 0.02869$\pm$0.00015 \citep{1999PASP..111..438F}. It was classified as a Type Ia-pec SN by \cite{2020TNSCR2512....1P}. Figure \ref{fig:ds9_2020rea} shows the location of SN~2020rea in UGC 10655. The details of SN~2020rea and its host galaxy are given in Table \ref{tab:SN2020rea_detail}.

Optical photometric follow-up of SN~2020rea was initiated $\sim$ 6 days after discovery with the telescopes of the Las Cumbres Observatory (LCO;  \citealt{2013PASP..125.1031B})
under the Global Supernova Project (GSP) in {\it BgVri} bands. SN~2020rea is located in the proximity of the host galaxy hence we performed template subtraction to estimate the true SN flux. The templates were observed in {\it BgVri} bands on May 27, 2021, $\sim$ 8 months after the discovery. The template subtraction was performed using \texttt{PyZOGY} \citep{2017zndo...1043973G}. The \texttt{lcogtsnpipe} pipeline  \citep{2016MNRAS.459.3939V} was used to estimate the SN magnitudes. Calibration of the instrumental magnitudes were done using APASS catalog. The calibrated photometric magnitudes of SN~2020rea are listed in Table \ref{tab:photometric_observational_log_2020rea}.

Spectroscopic follow up of SN~2020rea was initiated $\sim$ 5 days after discovery and lasted $\sim$ 1 month using the FLOYDS spectrograph on the 2m FTN telescopes. FLOYDS spectrograph provides a wavelength range of 3300-11000 \AA\ with resolution ranging between 400-700. We have used the \texttt{floydsspec}\footnote{https://www.authorea.com/users/598/articles/6566} pipeline to perform the spectral reduction. Finally the spectra were scaled with respect to the photometry and corrected for redshift. The log of spectroscopic observations is presented in Table \ref{tab:spectroscopic_observations_20rea}.

\section{Distance and extinction}
\label{distance_extinction}

\label{distance_extincton} Assuming $H_0$ = 73 km s$^{-1}$ Mpc$^{-1}$, $\Omega_m$ = 0.27,  $\Omega_v$ = 0.73 and a redshift of 0.02869$\pm$0.00015 we estimate the luminosity distance of SN~2020rea to be 120.5$\pm$6.7 Mpc. The distance modulus is 35.40 $\pm$ 0.12 mag. We quote the error from the HyperLeda database \citep{2014A&A...570A..13M}. The Galactic extinction along the line of sight in SN~2020rea is {\it E(B-V)} = 0.02 mag \citep{2011ApJ...737..103S}. SN~2020rea lies in the proximity of the host galaxy and hence extinction due to the host galaxy is also expected. To estimate the extinction due to the host galaxy we used the equivalent width of Na {\sc i}D line in the spectra. The initial spectral sequence of SN~2020rea shows the presence of a strong Na {\sc i}D line. We measured the equivalent width of Na {\sc i}D line in the spectrum combined using two spectra of SN~2020rea close to maximum (Figure \ref{fig:SN 2020rea_spectra_plot}). The estimated equivalent width is 0.66$\pm$0.06 \AA\ which translates to {\it E(B-V)} = $0.08\pm 0.02$ mag using the relation given in \cite{2012MNRAS.426.1465P}. Thus, the total extinction due to the combination of the Galactic and host components is {\it E(B-V)} = $0.10\pm0.02$ mag ({A${_V}$} = 0.31 mag assuming R${_V}$ = 3.1). 

\section{Analysis of the light curve}
\label{analaysis_light_curve}

\begin{figure}
	\begin{center}
		\includegraphics[width=\columnwidth]{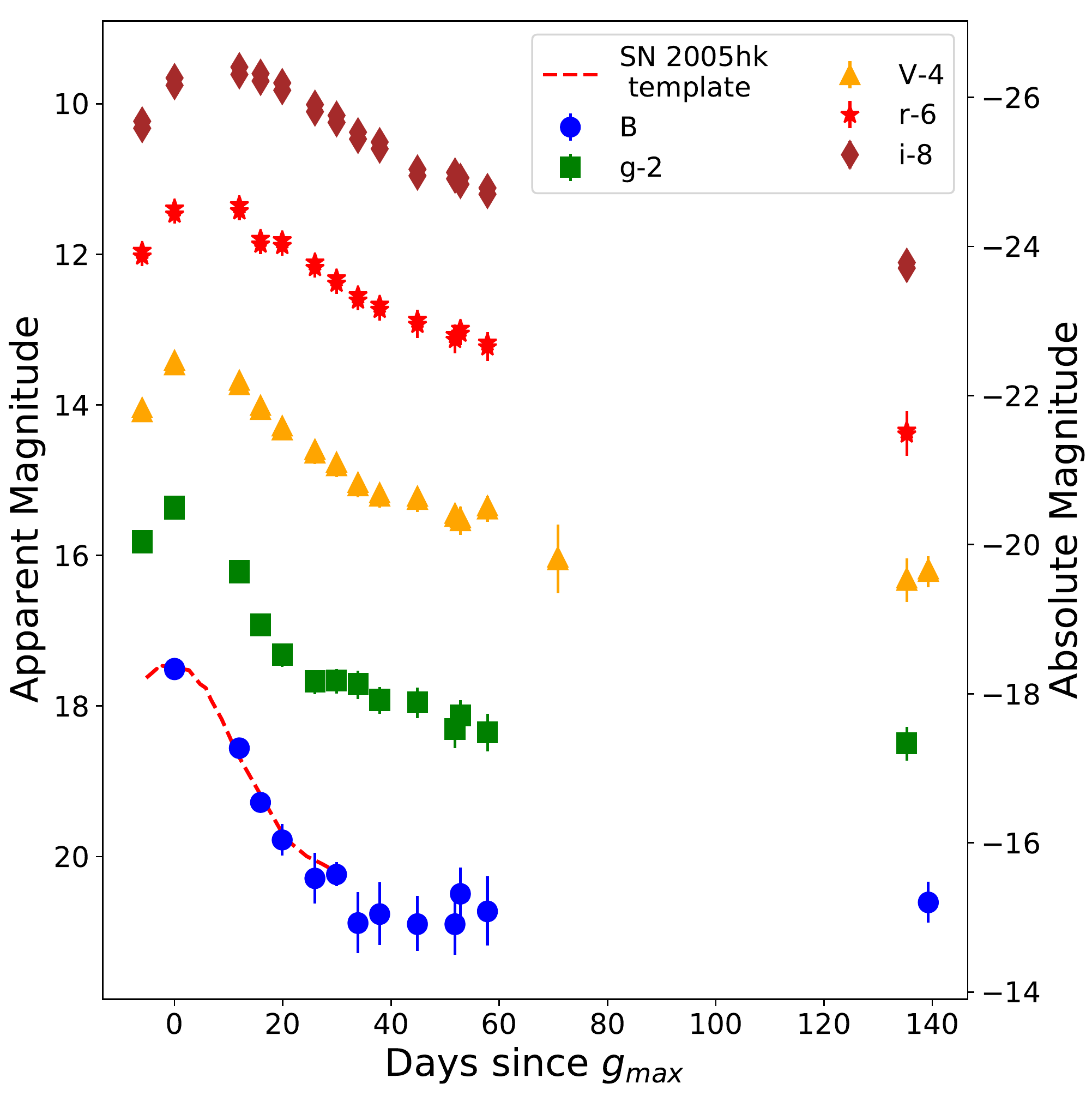}
	\end{center}
	\caption{Light curve evolution of SN~2020rea in {\it BgVri} bands. The light curves in all bands are shifted for clarity. In the right Y axis, corresponding absolute magnitudes for each band are presented. The template light curve of SN 2005hk used for estimating the peak magnitude and time of SN 2020rea in {\it B} band is also shown in the figure with dashed line.} 
	\label{fig:SN 2020rea_light_curve}
\end{figure}

Figure \ref{fig:SN 2020rea_light_curve} shows the light curve evolution of SN 2020rea in {\it BgVri} bands. The peak phase is well covered in all the bands except {\it B}-band. To  estimate the peak time and peak magnitude in {\it B}-band a chi-square minimization based template fitting method was used and a best match was found with SN 2005hk. The best fit indicates that SN~2020rea peaked at JD = 2459083.5$\pm$1 with peak magnitude 17.33$\pm$0.07 mag in the {\it B}-band . With these estimates, the light curve decline rate ($\Delta$m$_{15}$) of SN~2020rea in {\it B}-band is 1.61$\pm$0.14 mag. In other bands, peak phase and peak time are estimated by fitting a low order spline to the light curve.
The respective decline rates ($\Delta$m$_{15}$) in {\it g}, {\it V}, {\it r} and {\it i}-bands are 1.31$\pm$0.08 mag, 0.54$\pm$0.05 mag, 0.46$\pm$0.05 mag and 0.50$\pm$0.04 mag. The peak in  {\it g} and  V bands occur on JD = 2459084.74 and 2458084.77 at a magnitudes of 17.34$\pm$0.03 mag and 17.40$\pm$0.03 mag, respectively. We have used {\it g}-band maximum throughout the paper, as a reference, for further work. 

\begin{figure}
	\begin{center}
		\includegraphics[width=\columnwidth]{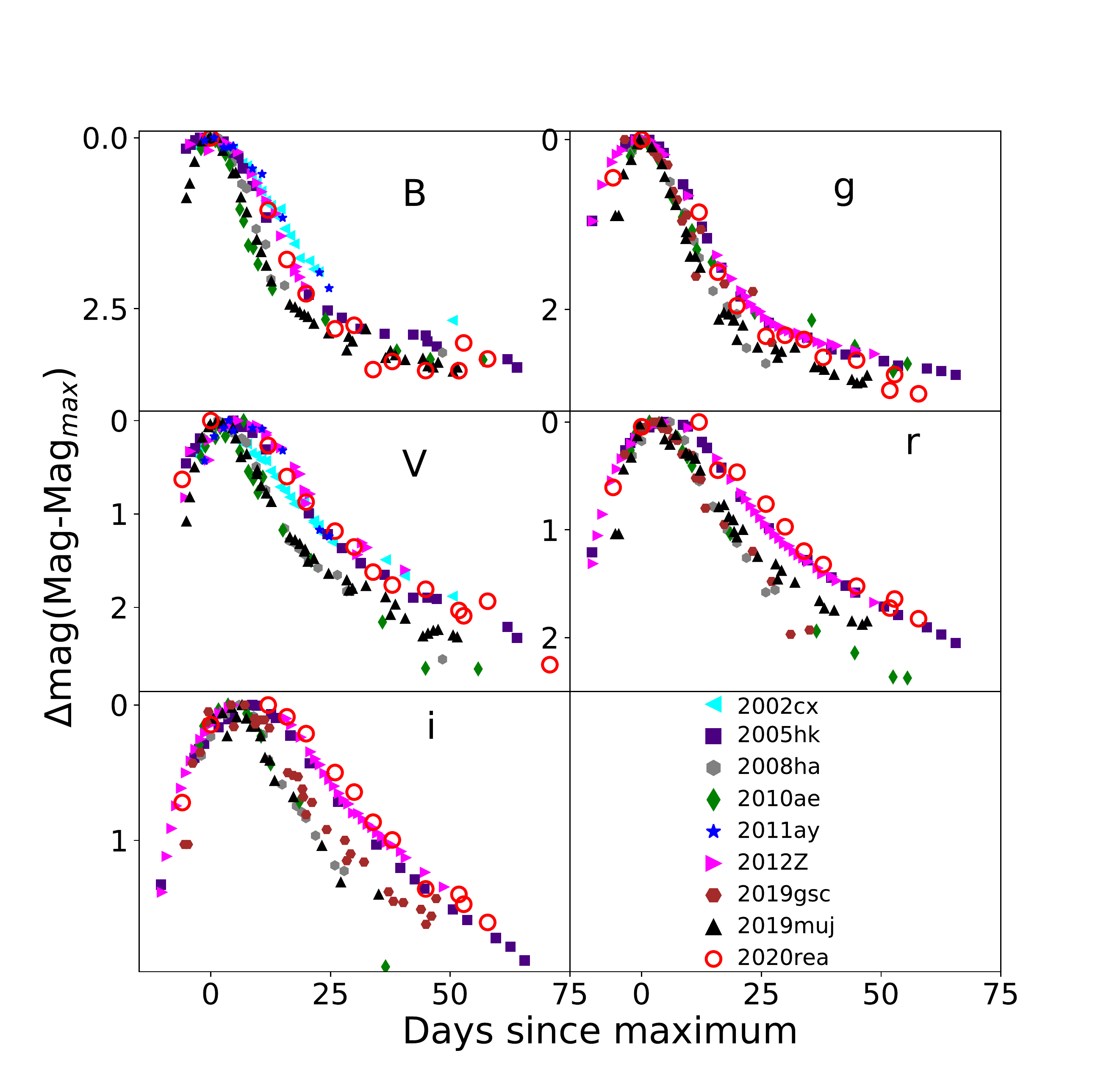}
	\end{center}
	\caption{Light curves of SN~2020rea in the {\it BgVri} bands and its comparison with other Type Iax SNe. Here, comparison plots in {\it B} and {\it V} bands are made with respect to maximum in {\it B} band while in {\it gri} bands comparison plots are constructed  with respect to {\it g} band maximum.}
	\label{fig:comp_light_curve_BgVri_2020rea}
\end{figure}

We compare the light curve characteristics of SN~2020rea with other well studied Type Iax SNe. We have represented the wide luminosity range in choosing the comparison sample which includes SNe 2002cx \citep{2003PASP..115..453L}, 2005hk \citep{2008ApJ...680..580S}, 2008ha \citep{2009AJ....138..376F}, 2010ae \citep{2014A&A...561A.146S}, 2011ay \citep{2015MNRAS.453.2103S}, 2012Z \citep{2015A&A...573A...2S,2015ApJ...806..191Y}, 2019muj \citep{2021MNRAS.501.1078B,10.1093/pasj/psab075} and 2019gsc \citep{2020ApJ...892L..24S}. Figures \ref{fig:comp_light_curve_BgVri_2020rea} exhibits  the normalized magnitudes of each SN with respect to the peak magnitude in the respective bands. In {\it B}-band, SN~2020rea declines faster than SNe 2002cx, 2011ay and follows a similar evolution as SNe 2005hk and 2012Z up to $\sim$ 20 days after maximum, whereas it declines faster than SN 2005hk at later epochs and shows similarity with SN 2019muj. In {\it V}-band, SN 2020rea shows resemblance with SNe 2005hk and 2012Z. The early time evolution of {\it g}-band light curve of SN~2020rea ($\Delta$m$_{15}$(g) = 1.31$\pm$0.08 mag) is similar to SNe 2005hk ($\Delta$m$_{15}$(g) = 1.36$\pm$0.01 mag, \cite{2015A&A...573A...2S})  and 2012Z ($\Delta$m$_{15}$(g) = 1.31$\pm$0.01 mag, \cite{2015A&A...573A...2S})  whereas in {\it r}-band SN 2020rea ($\Delta$m$_{15}$(r) = 0.46$\pm$0.05 mag) declines slightly slower than SNe 2005hk ($\Delta$m$_{15}$(r) = 0.70$\pm$0.02 mag, \cite{2015A&A...573A...2S}) and 2012Z ($\Delta$m$_{15}$(r) = 0.66$\pm$0.02 mag, \cite{2015A&A...573A...2S}) (Figure \ref{fig:comp_light_curve_BgVri_2020rea}). In {\it i}-band SN~2020rea ($\Delta$m$_{15}$(i) = 0.50$\pm$0.04 mag) shows similarity with SN 2012Z ($\Delta$m$_{15}$(i) = 0.54$\pm$0.04 mag, \citealt{2015A&A...573A...2S}) and declines slower than SN 2005hk ($\Delta$m$_{15}$(i) = 0.60$\pm$0.01 mag, \citealt{2015A&A...573A...2S}).

\begin{figure}
	\begin{center}
		\includegraphics[width=\columnwidth]{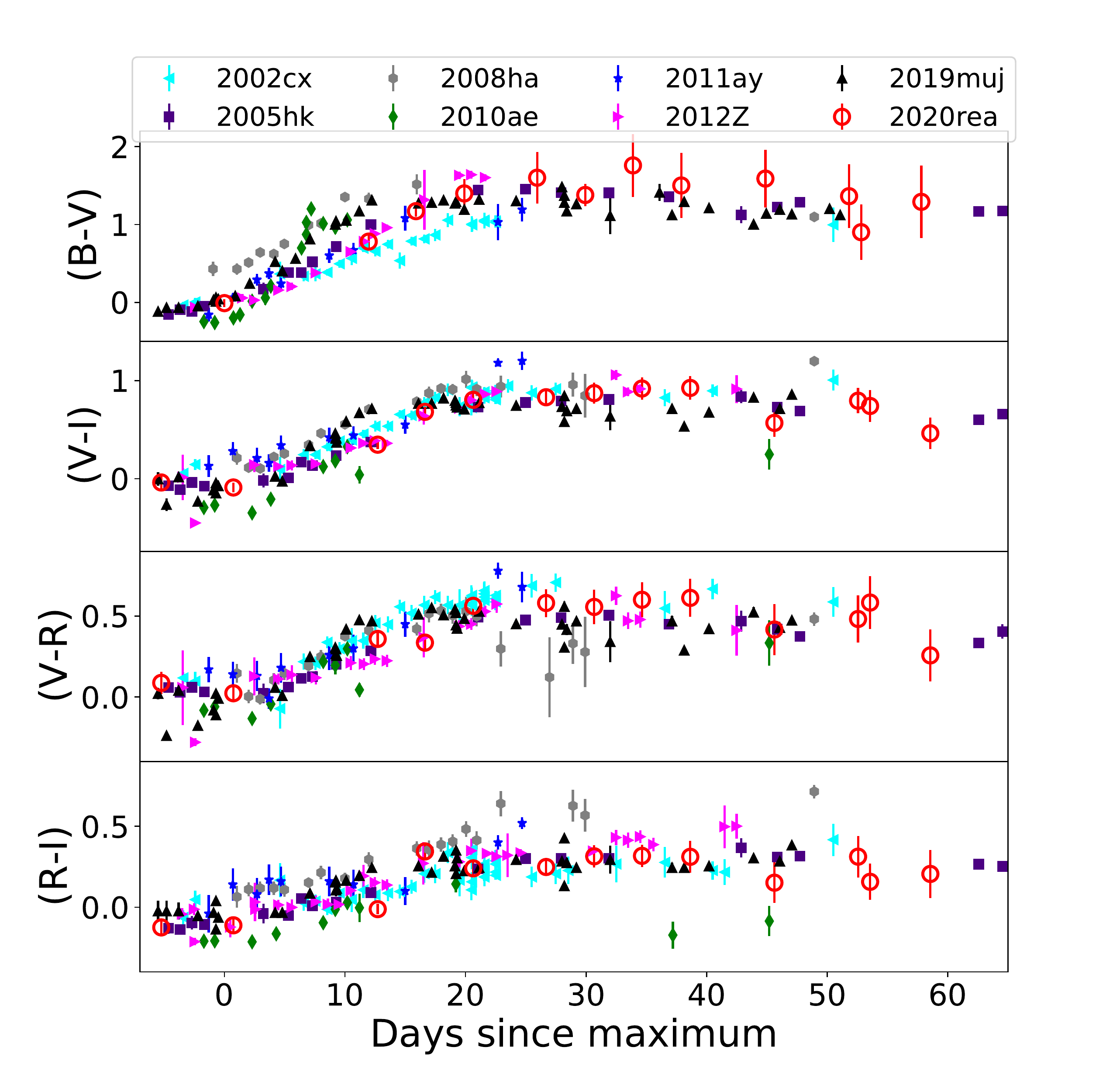}
	\end{center}
	\caption{The colour evolution of SN~2020rea and its comparison with colours of other well studied Type Iax SNe.} 
	\label{fig:SN 2020rea_colour_curve}
\end{figure}

Figure \ref{fig:SN 2020rea_colour_curve} presents reddening corrected {\it (B-V)}, {\it (V-I)}, {\it (V-R)} and {\it (R-I)} colour evolution of SN~2020rea and its comparison with other Type Iax SNe. For SNe 2020rea and 2010ae, we have used the formulations given in \cite{2006A&A...460..339J} to convert {\it ri} magnitude into {\it RI} magnitude. The {\it (B-V)}, {\it (V-I)}, {\it (V-R)} and {\it (R-I)} colour evolution of SN~2020rea follows a trend similar to other Type Iax SNe used for comparison. We have used date of {\it B} band maximum as reference for SNe 2002cx and 2011ay and {\it g} band maximum as reference for all the other SNe used for comparison.
 
Using the distance and extinction given in Section \ref{distance_extinction}, we estimate the peak absolute magnitude of SN 2020rea in {\it V}-band = $-$18.30$\pm$0.12 mag. This is similar to SNe 2011ay \citep{2015A&A...573A...2S}, 2012Z \citep{2015A&A...573A...2S} and higher than SNe 2002cx \citep{2003PASP..115..453L}, 2005hk \citep{2008ApJ...680..580S} and 2014dt \citep{2018MNRAS.474.2551S}. Absolute magnitudes of SN~2020rea in {\it BgVri} bands are presented in Figure \ref{fig:SN 2020rea_light_curve}.   

\section{Light curve modelling}
\label{bolometric_light_curve}

We construct the pseudo-bolometric light curve of SN~2020rea using extinction corrected magnitudes in {\it BgVri} bands. For the epoch JD~2459078.8, the $B$-band photometry is missing and hence we fit the $B$-band light curve with the template of SN~2005hk \citep{2008ApJ...680..580S} to estimate the magnitude. The extinction corrected magnitudes were converted to flux using zero points from the SVO filter profile service \footnote{\url{http://svo2.cab.inta-csic.es/theory/fps/index.php?mode=browse&gname=LCO&asttype=}} \citep{2020sea..confE.182R}. These fluxes are used to generate spectral energy distribution (SED) at each epoch which was then integrated using trapezoidal rule between 4000 to 9000 \AA~ to get the pseudo-bolometric flux. The contribution of UV and IR flux to the total bolometric flux is not well constrained for Type Iax SNe. It is estimated to be typically lying in the range 10\% to 53\% \citep{2007PASP..119..360P,2015ApJ...806..191Y,2016MNRAS.459.1018T,2020MNRAS.496.1132T,2020ApJ...892L..24S,2022ApJ...925..217D}. Due to unavailability of data in UV and IR bands, we have used pseudo-bolometric fluxes and reported the lower limit of the explosion parameters.

\begin{figure}
	\begin{center}
	    \includegraphics[width=\columnwidth]{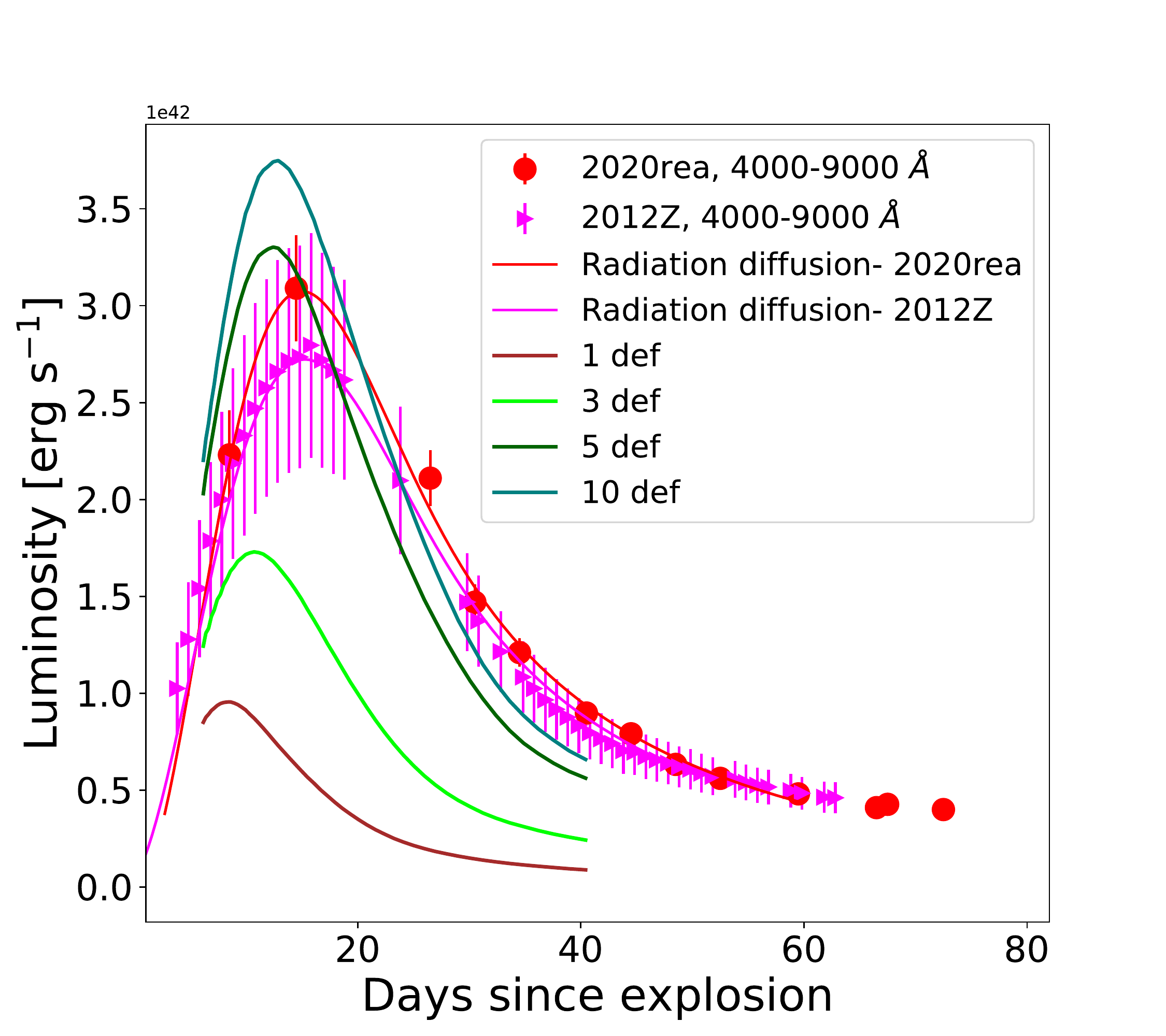}
	\end{center}
	\caption{Pseudo-bolometric light curves of SNe~2020rea and 2012Z fitted with the radiation diffusion model are shown. The pseudo-bolometric light curves are compared with optical bolometric light curves of the pure deflagration of $M_{\rm ch}$ white dwarf \citep{2014MNRAS.438.1762F}.}
	\label{fig:2020rea_2012Z_deflag_bol_light_curve}
\end{figure}

The integrated fluxes are converted to luminosity using the distance modulus $\mu$ = 35.40 $\pm$ 0.12 mag. The peak pseudo-bolometric luminosity of SN~2020rea is (3.09 $\pm$ 0.27) $\times$ 10$^{42}$ erg s$^{-1}$ and it occurred at JD 2459087.26 about 2.52 days after maximum in $g$-band. For direct comparison, we also estimate the pseudo-bolometric light curve of SN~2012Z using {\it BgVri} data with $E(B-V)$ = 0.11 $\pm$ 0.03 mag \citep{2015A&A...573A...2S} and distance modulus of 32.34 $\pm$ 0.28 mag, obtained using the luminosity distance of 29.4$\pm$3.8 Mpc. The peak pseudo-bolometric luminosity of SN~2012Z is (2.82 $\pm$ 0.58) $\times$ 10$^{42}$ erg s$^{-1}$ at JD 2455972.0. The peak pseudo-bolometric luminosity of SN~2020rea is slightly higher than SN 2012Z and lies towards the brighter end of the luminosity distribution of Type Iax SNe. Figure \ref{fig:2020rea_2012Z_deflag_bol_light_curve} shows the pseudo-bolometric light curves of SNe~2020rea and 2012Z.

To constrain the amount of $^{56}$Ni synthesized during the explosion we used a radiation diffusion model (\citealt{1982ApJ...253..785A, 2008MNRAS.383.1485V,2012ApJ...746..121C}) which takes into account energy generated through radioactive decay from $^{56}$Ni $\rightarrow$  $^{56}$Co $\rightarrow$ $^{56}$Fe and also includes $\gamma$-ray escape from the ejecta.

The output luminosity is expressed as 
\begin{equation}
\begin{split}
\label{eq:Arnett}
    L(t) = M_{\rm{Ni}} \mathrm{e}^{-x^{2}} [(\epsilon_{\rm{Ni}}-\epsilon_{\rm{Co}})\int_0^x 2 z \mathrm{e}^{z^{2}-2zy}\,\mathrm{d}z\\
         + \epsilon_{\rm{Co}}\int_0^x 2 z \mathrm{e}^{z^{2}-2yz+2zs}\,\mathrm{d}z]( 1 - \mathrm{e}^{-{(\frac{t_{\gamma}}{t}})^{2}}) 
\end{split}  
\end{equation}

\noindent
where $t$ (days) is the time since explosion, \(t_{\rm{lc}}\) is time scale of the light curve, $t_\gamma$ is gamma ray time scale, \(M_{\rm{Ni}}\) is initial mass of $^{56}$Ni, $x$ $\equiv$ $t$/$t$\(_{\rm{lc}}\),  $y$ $\equiv$ \(t_{\rm{lc}}\)/(2\(t_{\rm{Ni}}\)) and $s$ $\equiv$ [\(t_{\rm{lc}}\)(\(t_{\rm{Co}}\) - \(t_{\rm{Ni}}\))/(2\(t_{\rm{Co}}\)\(t_{\rm{Ni}}\))] with \(t_{\rm{Ni}}\) = 8.8~d and \(t_{\rm{Co}}\) = 111.3~d, respectively. The rate of energy generation due to Ni and Co decay are  \(\rm \epsilon_{Ni} = 3.9 \times 10^{10}\ erg\ s^{-1}\ g^{-1}\) and \(\rm \epsilon_{Co} = 6.8 \times 10^{9}\ erg\ s^{-1}\ g^{-1}\), respectively. The free parameters in the model are epoch of explosion $t_{expl}$, $M_{\rm{Ni}}$, $t_\gamma$ and $t_{\rm{lc}}$.  

The mass of ejecta (\(M_{\rm{ej}}\)) and kinetic energy (\(E_{\rm{K}}\)) are expressed as 

\begin{equation}
\label{eq:EjectaMass}
    M_{\rm{ej}} = 0.5 \frac{\beta c}{\kappa} v_{exp}t_{lc}^{2} 
\end{equation}

\begin{equation}
\label{eq:KineticEnergy}
    E_{\rm{K}} = 0.3 M_{ej} v_{exp}^{2}
\end{equation} 

\noindent
where $v_{exp}$, $c$ and $\beta$ (= 13.8) are the expansion velocity of the ejecta, the speed of light, and the constant of integration, respectively. 

The fit of the radiation diffusion model to the pseudo-bolometric light curve of SN~2020rea gives $^{56}$Ni = 0.13$^{+0.01}_{-0.01}$ M$_{\odot}$, $t_{\rm lc}$ = 12.36$^{+0.9}_{-1.75}$ days, $t_{\rm \gamma}$ = 43.60$^{+2.4}_{-1.7}$ days and $JD_{\rm exp}$ = 2459070.64$^{+1.45}_{-0.76}$. The ejecta mass for SN 2020rea is estimated as $M_{\rm ej}$ = 0.77$^{+0.11}_{-0.21}$ M$_{\odot}$ and kinetic energy $KE$ = 0.19$^{+0.02}_{-0.06}$ $\times$ 10$^{51}$ erg, using a constant opacity $\kappa_{\rm opt}$ = 0.1 cm$^{2}$g$^{-1}$ and $v_{\rm exp}$ of 6500 km s$^{-1}$, close to maximum light.

We repeat the same exercise for the pseudo-bolometric light curve of SN~2012Z. We get $^{56}$Ni = 0.12$^{+0.01}_{-0.01}$, $t_{\rm lc}$ = 14.19$^{+0.8}_{-1.2}$ days, $t_{\rm \gamma}$ = 43.68$^{+1.2}_{-1.5}$ days and $JD_{\rm exp}$ = 2455954.39$^{+0.5}_{-0.37}$. Using an expansion velocity of 7000 km~s$^{-1}$ and the same constant optical opacity, we get $M_{\rm ej}$ = 1.09$^{+0.12}_{-0.19}$ M$_{\odot}$ and $KE$ = 0.32$^{+0.04}_{-0.05}$ $\times$ 10$^{51}$ erg. The values of $^{56}$Ni mass, ejecta mass and kinetic energy estimated by \cite{2014A&A...561A.146S} for SN 2012Z are 0.25--0.29 $M_{\odot}$, 1.4--2.6 $M_{\odot}$ and 0.7--2.8 $\times$ 10$^{51}$ erg, respectively which are higher than our estimates. The difference is mostly due to the adopted distance modulus, the wavelength range of the spectral energy distribution and velocity used for estimating the explosion parameters.  The faster rise in SN~2020rea as compared to SN 2012Z could be attributed to the different amount of $^{56}$Ni mixing in the ejecta. 

We compare the pseudo-bolometric light curves of SN~2020rea and SN~2012Z with optical bolometric light curves of pure deflagration model of $M_{\rm ch}$ white dwarfs \citep{2014MNRAS.438.1762F}. For each model mentioned in Figure \ref{fig:2020rea_2012Z_deflag_bol_light_curve}, we integrate the model optical spectrum at each epoch available with the \texttt{HESMA} database in the same wavelength range as for SN~2020rea to obtain the model pseudo-bolometric luminosity. In the deflagration models, the explosion strength is characterized by ignition spots. With the increase in number of ignition spots, more material burns, which leads to an increase in the luminosity, explosion energy and ejecta velocity. The model light curves for N1-def, N3-def, N5-def and N10-def, with ignition spots 1, 3, 5, 10, respectively, are shown in Figure \ref{fig:2020rea_2012Z_deflag_bol_light_curve}. 

The early photospheric phase of the light curve for SN 2020rea falls between models N3-def and N5-def. However, the observed light curves of both SNe~2012Z and 2020rea declines slower than the N5-def as well as the N10-def model bolometric light curves. This is because the ejected mass, the parameter that accounts for the decline rate, in the N5-def and N10-def models are 0.372 and 0.478 M$_{\odot}$, respectively \citep{2014MNRAS.438.1762F}, which are less than the estimated ejecta mass for SNe 2012Z and 2020rea.

\section{spectral properties}
\label{spectral_properties}

\begin{figure}
	\begin{center}
		\includegraphics[width=\columnwidth]{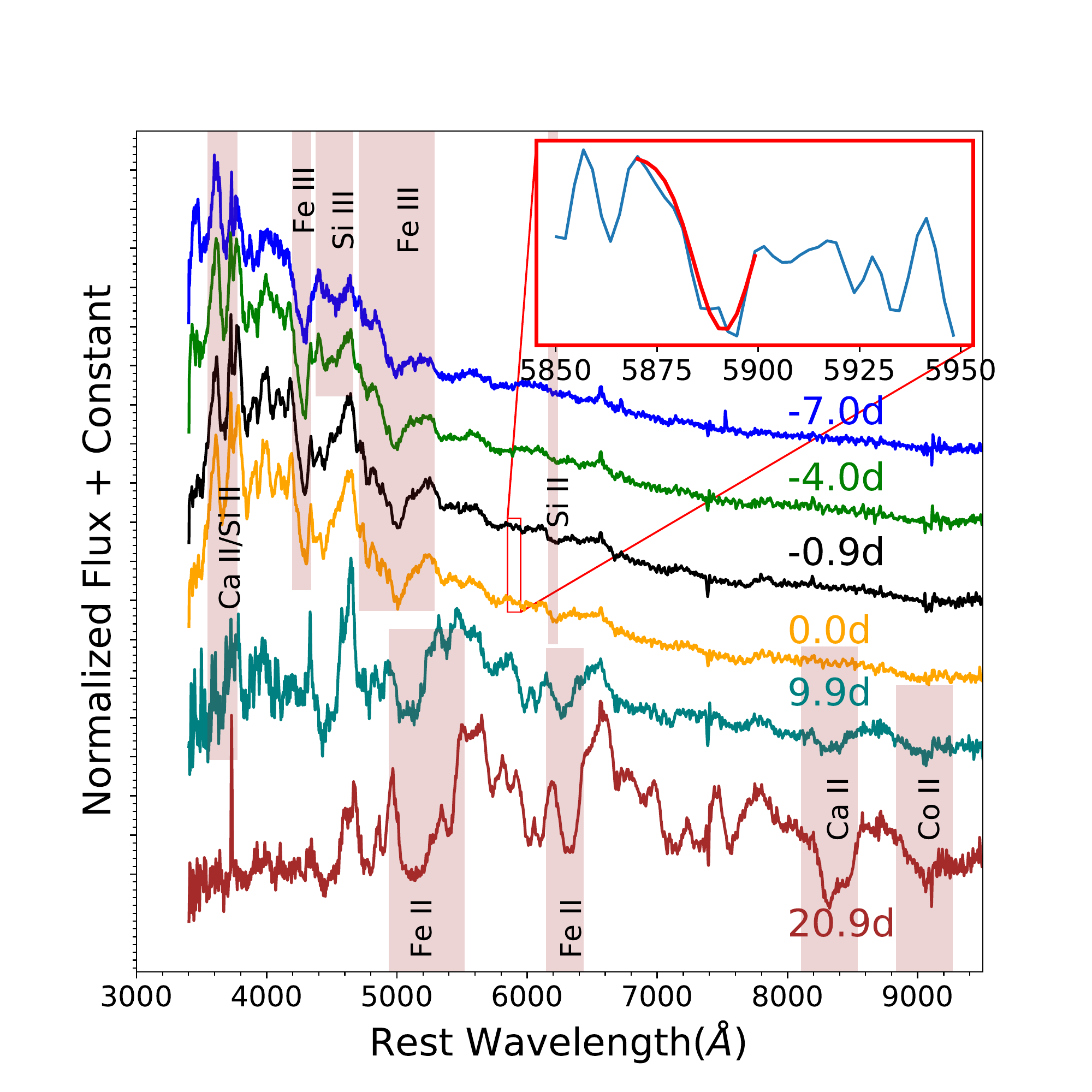}
	\end{center}
	\caption{Spectral evolution of SN~2020rea spanning between $-$7.0 days to +20.9 days since maximum in {\it g}-band. Prominent lines are marked with shaded bars. In the inset plot, we have presented the zoomed Na {\sc i}D feature and the fit associated with it.}
	\label{fig:SN 2020rea_spectra_plot}
\end{figure}

Figure \ref{fig:SN 2020rea_spectra_plot} presents the spectral evolution of SN 2020rea from $\sim$ $-$7 days to +21 days. The early time spectra are dominated by a blue continuum along with well developed P-Cygni profiles with relatively broad absorption features. The pre-maximum spectra of SN 2020rea show Si {\sc II}/Ca {\sc II} feature in the blue region, Fe {\sc III}, Si {\sc III}, S {\sc II} and relatively weak Si {\sc II} feature around 6000 \AA. The spectrum around maximum is similar to the pre-maximum spectra with an evolved Si {\sc II} feature. After maximum, a feature at $\sim$ 6000 \AA\ grows stronger and can be associated with Fe {\sc II}. In the 8000 \AA\ to 9000 \AA\ region, the Ca II NIR triplet starts developing. A clear absorption feature due to Co {\sc II} $\sim$ 9000 \AA\ is also present. The spectral region between 5500 \AA\ and 7000 \AA\ is dominated by Fe {\sc II} lines. By +21 days the continuum becomes redder and Co {\sc II} around 6600 \AA\ starts developing. In addition, Fe {\sc II} feature in the blue region, Ca {\sc II} NIR triplet and Co {\sc II} at $\sim$ 9000 \AA\ become stronger. 

\subsection{Comparison with other Type Iax SNe}
\label{comparion_spectral_features_other_SNe}

\begin{figure}
	\begin{center}
		\includegraphics[width=\columnwidth]{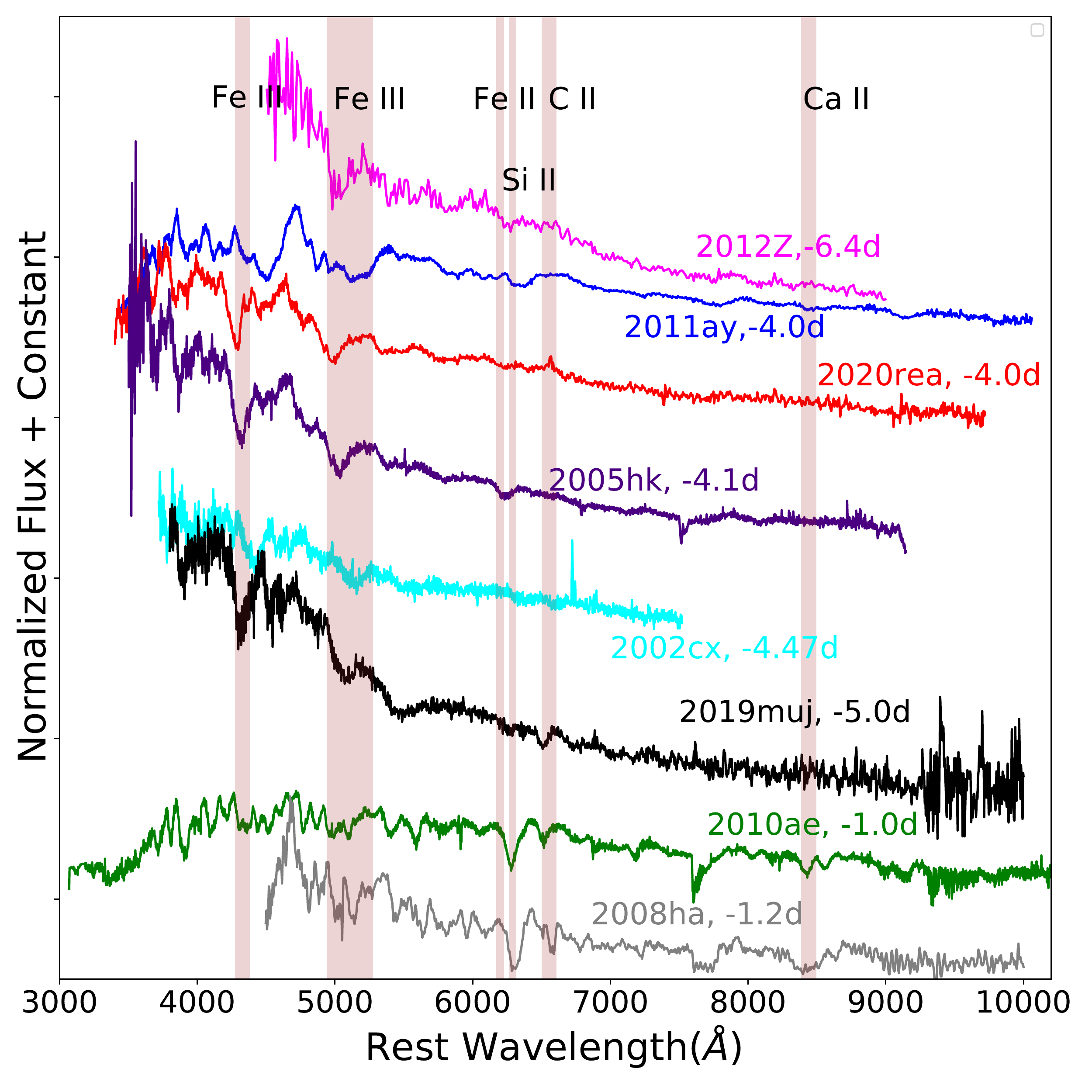}
	\end{center}
	\caption{Comparison of pre-maximum spectrum of SN~2020rea with other well studied Type Iax SNe.}
	\label{fig:SN 2020rea_spectra_comp_pre_peak}
\end{figure}

\begin{figure}
	\begin{center}
		\includegraphics[width=\columnwidth]{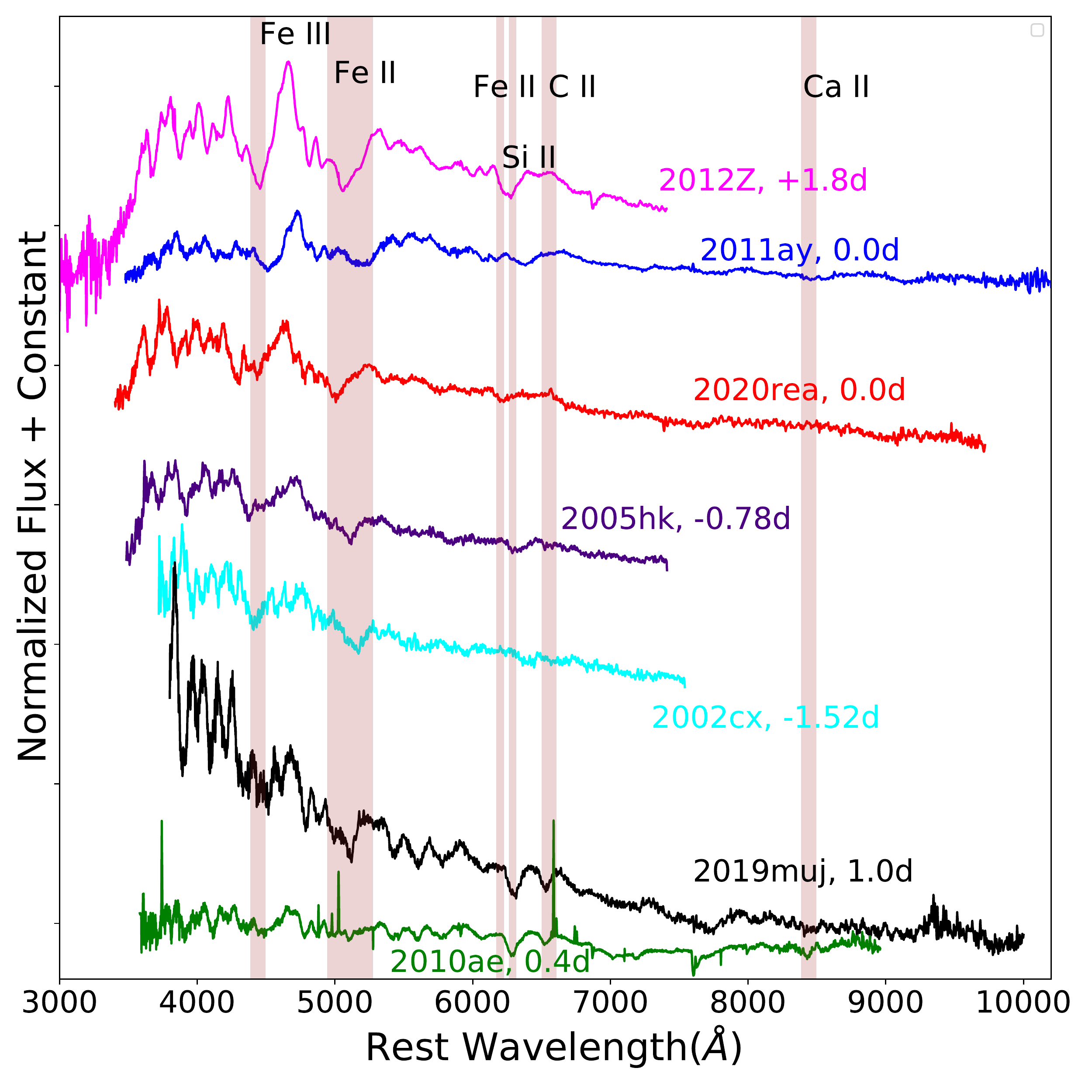}
	\end{center}
	\caption{Near maximum spectrum of SN~2020rea is shown with spectra of other Type Iax SNe at comparable epochs.}
	\label{fig:SN 2020rea_spectra_comp_peak}
\end{figure}

\begin{figure}
	\begin{center}
		\includegraphics[width=\columnwidth]{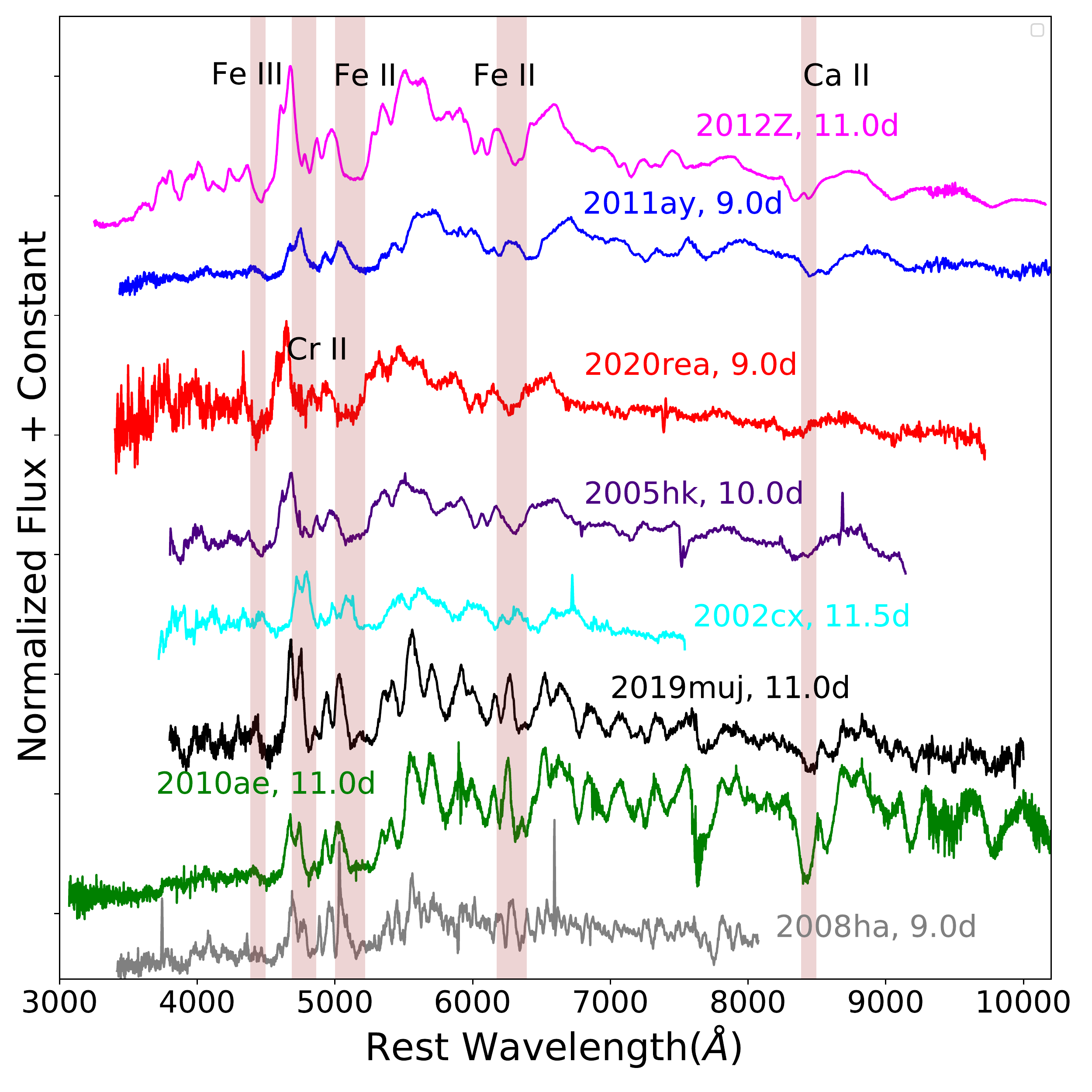}
	\end{center}
	\caption{The post-maximum spectrum of 
	SN~2020rea compared with spectra of other Type Iax SNe at similar epoch.}
	\label{fig:SN 2020rea_spectra_comp_post_peak}
\end{figure}

\begin{figure}
	\begin{center}
		\includegraphics[width=\columnwidth]{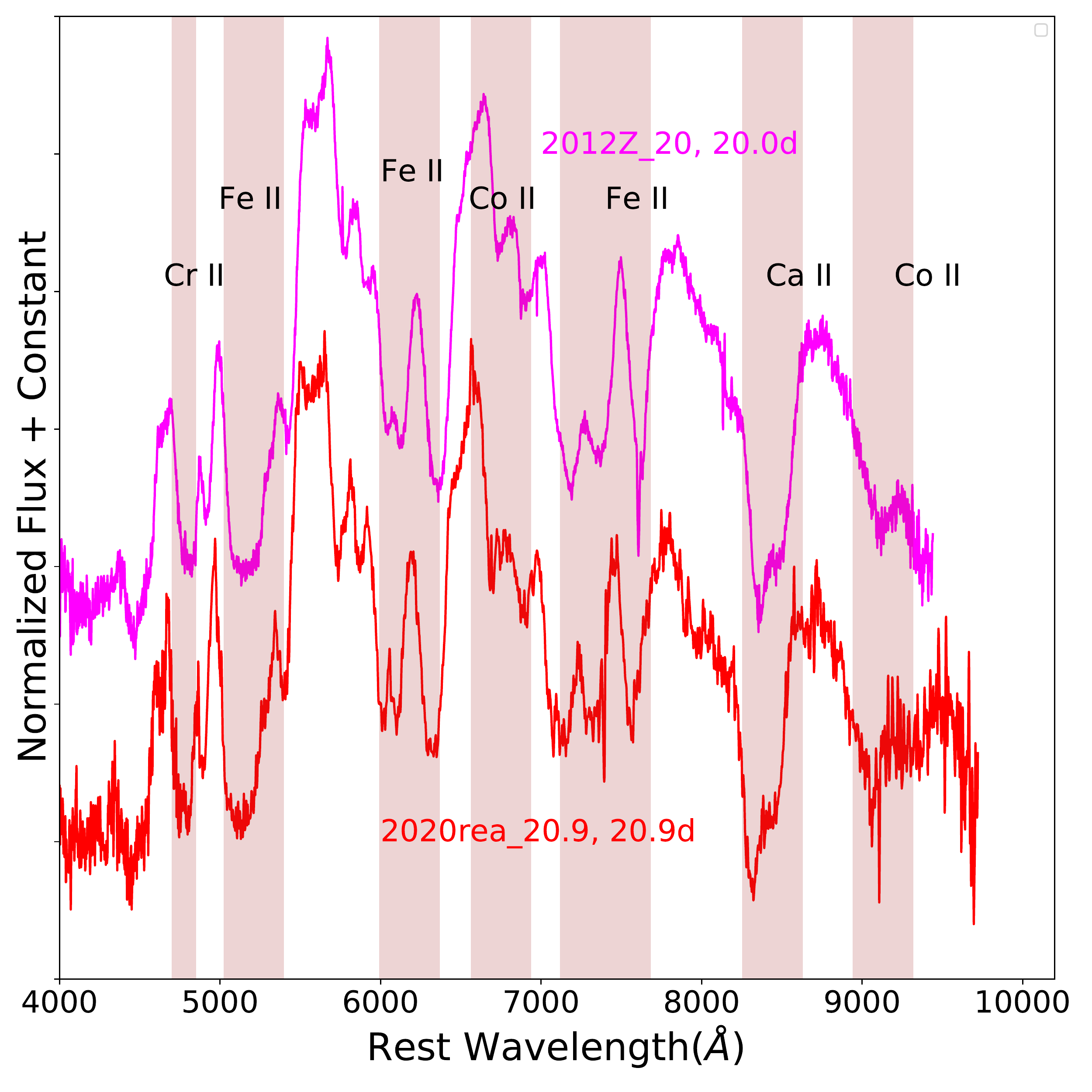}
	\end{center}
	\caption{Comparison of spectral features of SN~2020rea at +21 day with SN 2012Z.}
	\label{fig:SN 2020rea_spectra_comp_21_day}
\end{figure}

To investigate the nature of spectral lines we compare the pre-maximum, near maximum and post-maximum spectra of SN 2020rea with other well studied Type Iax SNe such as SNe 2002cx \citep{2003PASP..115..453L}, 2005hk \citep{2007PASP..119..360P,2008ApJ...680..580S}, 2008ha \citep{2009Natur.459..674V,2009AJ....138..376F}, 2010ae \citep{2014A&A...561A.146S}, 2011ay \citep{2013ApJ...767...57F}, 2012Z \citep{2013ApJ...767...57F,2015A&A...573A...2S} and 2019muj \citep{2021MNRAS.501.1078B}.
Figure \ref{fig:SN 2020rea_spectra_comp_pre_peak} presents the pre-maximum spectra of SN 2020rea and other Type Iax SNe. The Fe {\sc III} feature near 4000 \AA\ and 5000 \AA\ are seen in all the SNe having coverage in bluer region. The C {\sc II} feature is prominent in fainter and intermediate luminosity Type Iax SNe 2008ha, 2010ae and 2019muj, however, in SN 2020rea and other bright Type Iax SNe, this feature is very weak. The Ca {\sc II} NIR triplet can only be seen in SNe 2008ha and 2010ae. Overall pre-maximum spectroscopic features of SN~2020rea are typical of brighter Type Iax SN. In the spectral comparison near maximum, we find that the prominent spectral lines such as Fe {\sc III}, Fe {\sc II} and Si {\sc II} are present in all the SNe as shown in Figure \ref{fig:SN 2020rea_spectra_comp_peak}. In the post maximum spectra (Figure \ref{fig:SN 2020rea_spectra_comp_post_peak}), the Ca {\sc II} NIR feature is clearly seen in SNe 2005hk, 2010ae, 2011ay, 2012Z and 2019muj. SN~2020rea has weak Ca {\sc II} NIR triplet. The Fe {\sc III}, Fe {\sc II} multiplets and Cr {\sc II} lines are clearly visible in all the SNe. At the post maximum phase, SNe 2020rea and 2012Z show resemblance in their spectral properties. For a detailed spectral comparison between SNe 2012Z and 2020rea, spectra obtained $\sim$ 20 days after maximum of both the SNe are plotted in Figure \ref{fig:SN 2020rea_spectra_comp_21_day}. We notice that both the SNe show similarities with each other in terms of spectral signatures, displaying relatively broad features. 

\begin{figure}
	\begin{center}
		\includegraphics[width=\columnwidth]{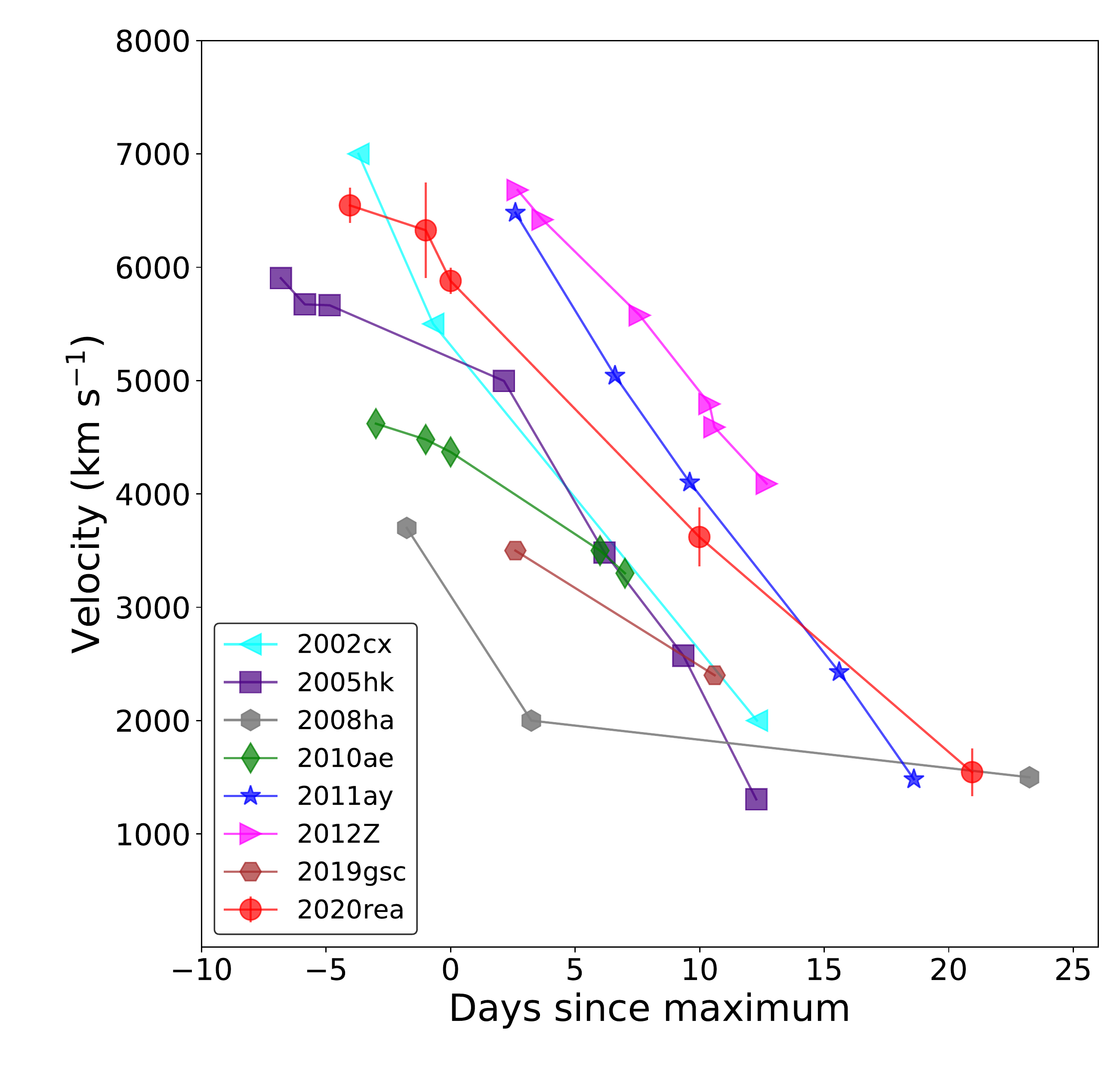}
	\end{center}
	\caption{Velocity evolution of Si {\sc II} line of SN~2020rea and its comparison with other well studied Type Iax SNe. Error bars associated with velocity estimation of SN~2020rea are also plotted in the figure.}
	\label{fig:velocity_plot}
\end{figure}

Figure \ref{fig:velocity_plot} shows the velocity evolution of the Si {\sc II} 6355 \AA\ feature of SN~2020rea and other Type Iax SNe. The line velocities are measured by fitting Gaussian profiles to the absorption minima of the P-Cygni profile associated with Si {\sc II} line. The error bar associated with velocities of SN~2020rea are measurement errors only. In the pre-maximum phase, the line velocity of the Si {\sc II} feature in SN~2020rea is less than SN 2002cx and higher than SN 2005hk. In the post-maximum phases the Si {\sc II} line velocity of SN~2020rea is lower than SNe 2011ay, 2012Z and higher than other comparison SNe. In the late post-maximum phase, the identification of Si {\sc II} is a bit questionable as Fe {\sc II} lines (at 6149 \AA\ and 6247 \AA) start appearing close to the Si {\sc II} line.

The velocity of the Fe {\sc II} 5156 \AA\ line in the pre-maximum and near maximum spectra are estimated as $\sim$ 10000 km s$^{-1}$ and 8570 km s$^{-1}$, respectively which are around 3500 km s$^{-1}$ and 2000 km s$^{-1}$ higher than the Si {\sc II} velocity at similar phase. This trend of higher velocity of Fe {\sc II} lines as compared to Si {\sc II} line shows significant mixing of burned materials \citep{2007PASP..119..360P}. 

\subsection{Spectral modelling}
\label{spectral_modelling}

We perform modelling of a few spectra of SN~2020rea using \texttt{TARDIS} (a one dimensional radiative transfer code,  \citealt{2014MNRAS.440..387K,kerzendorf_wolfgang_2018_1292315}). \texttt{TARDIS} assumes an opaque core with a sharp boundary or photosphere that emits a blackbody continuum. The ejecta is divided into spherical shells and is assumed to be undergoing homologous expansion. \texttt{TARDIS} allows the user to supply custom density and abundance profiles for the SN ejecta as input. In this work, we assume a uniform abundance profile for each element. The other input parameters are time since explosion and luminosity at a comparable epoch of the spectrum. The photospheric approximation used in \texttt{TARDIS} means that it is only applicable at early times. To generate the synthetic spectrum, we use as input the bolometric luminosity at the corresponding epoch. The mass fractions of radioactive isotopes are varied to improve the fit. For SN ejecta we adopt an exponential density profile of the form 

\noindent

\begin{equation}
    \rho(v,t_{exp}) = \rho_{0}(\frac{t_{0}}{t_{exp}})^{3}e^{-v/v_{0}}
\end{equation}

\noindent
where $t_{0}$ = 2 days, $\rho$$_{0}$ is reference density (= 6$\times$10$^{-11}$ g cm$^{-3}$), $t_{exp}$ is time since explosion, $v$ is velocity and $v$$_{0}$ is the reference velocity.

\begin{figure}
	\begin{center}
		\includegraphics[width=\columnwidth]{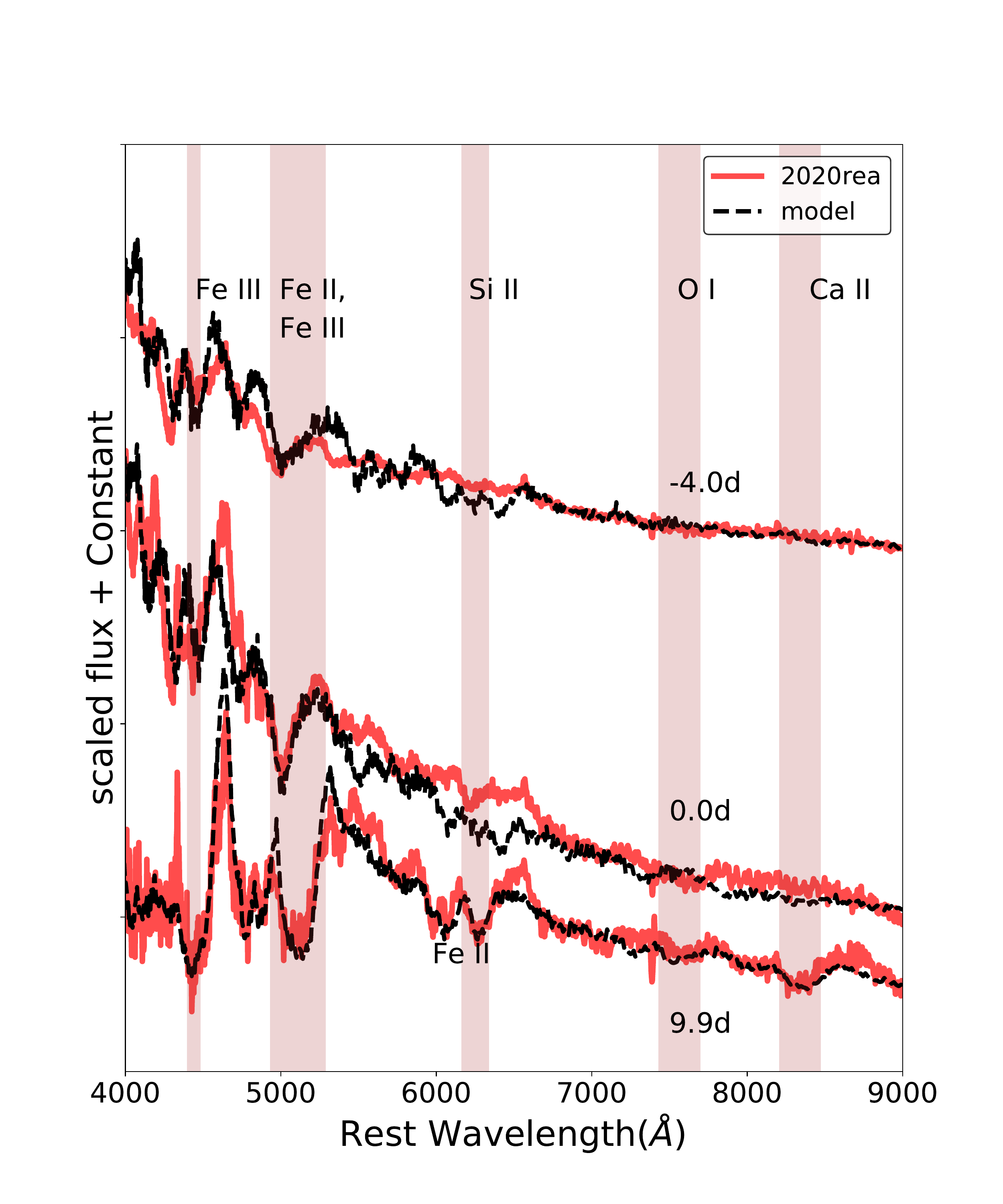}
	\end{center}
	\caption{Spectra of SN~2020rea during the photospheric phase, overplotted are the model spectra generated using TARDIS.}
	\label{fig:tardis_plot}
\end{figure}

In order to perform the \texttt{TARDIS} spectral fitting we adopt $v{_0}$ = 7000 km s$^{-1}$ and explosion time $t_{exp}$ JD = 2459070 (see section \ref{bolometric_light_curve} for details). The outer velocity of the ejecta has been fixed at 11500 km s$^{-1}$ and the inner velocity was varied between 6800 and 6000 km s$^{-1}$. Since there is degeneracy in the parameters used in \texttt{TARDIS} fit, the spectral model presented in this paper is not unique. The modelled spectra for $-$4.0, 0.0 and +9.9 days with respect to {\it g}-band maximum are overplotted on the observed spectrum in Figure \ref{fig:tardis_plot}. To model the observed spectra, species of carbon, oxygen, iron, cobalt, calcium, chromium, titanium and other ions usually present in SN ejecta are used. As we did not detect lines due to helium in the spectra, helium is not included in the model. 

\begin{table*}
\caption{ Parameters used in TARDIS model for SN 2020rea  }
\begin{tabular}{l  c c c c c c c c c c c c c c}
\hline \hline
t$\dagger$       & L                   & v$_{inner}$         & X(Si)         & X(C)       & X(O)   & X(S)   & X(Ni) & X(Ca)  & X(Co)    & X(Fe)   &  X(Ne) & X(Mg) & X(Cr) & X(Ti)     \\
(days)         & (logL/L$_{\odot}$)   &  kms$^{-1}$          																			                                \\
\hline
-4.0           &  8.9				 &   6800              &  0.02         &  0.02      &  0.10  & 0.004 & 0.4   &  0.003 &  0.006    & 0.00005 & 0.20695   & 0.04   & 0.18  & 0.01           \\
0.0            &  9.15               &   6500              &  0.02         &  0.02      &  0.10  & 0.004 & 0.4   &  0.003 &  0.006    & 0.0005	& 0.2065   & 0.04   & 0.18  & 0.01            \\
9.9            &  8.99               &   6000              &  0.02         &  0.02      &  0.10  & 0.004 & 0.4   &  0.003 &  0.005    & 0.35    & 0.018   & 0.04   & 0.02  & 0.01          \\
\hline\hline
\end{tabular}
\newline
$^\dagger$with respect to g$_{max}$= 2459084.74
\label{tab:tardis_20rea}      
\end{table*}

Table \ref{tab:tardis_20rea} presents the mass fraction of the dominant elements used to generate the model spectra (Figure \ref{fig:tardis_plot}). In the modelled spectrum at $-$4.0 day, Fe features between 4000 \AA\ and 5000 \AA\ are well reproduced, Si {\sc II} line is weak and continuum matches well with the observed spectrum. To constrain  mass fraction of Si, synthetic spectra were generated  by varying Si mass fraction at different epochs. It is found that increasing Si mass fraction beyond 1\% for pre-peak spectrum and 3\% for post-peak spectrum degrades the fit. Hence, we have used 2\% of Si for spectral fitting at all the three epochs. We do not see strong features due to C and O in the spectra, usually they are used as filler elements. However, we do see a weak OI line in the spectrum obtained at maximum and +9.9 d. We have used a significant amount of Ni for fitting all three spectra of SN~2020rea presented in Figure \ref{fig:tardis_plot}. In the synthetic spectra at pre-maximum and at maximum a very low amount of Fe is used as introducing more Fe resulted in over represented Fe features. We have included $\sim$ 20\% Neon as a filler element for fitting the first two epochs and $\sim$ 2\% of Ne for fitting the last spectrum at +9.9 day since maximum. IMEs such as Mg, Ca, S etc. are also used to fit the spectra. In the modelled spectrum around maximum, the region between 4000 \AA\ to 5200 \AA\ is similar to the observed spectrum. In the +9.9 day spectrum, the observed spectral features and continuum are well reproduced by the model with significant amount of IGEs. However, the `W' feature at $\sim$ 6000 \AA\ could not be reproduced. This feature is attributed to the presence of S line during the early phase of evolution which is later converted to iron when the SN enters the Fe dominated phase. Since we have assumed a model with a uniform abundance profile for each element and got a fairly good fit for our +9.9 day spectrum, this indicates towards a well mixed ejecta, which is expected in a deflagration scenario \citep{2003Sci...299...77G}.

\subsection{Host galaxy metallicity}
\label{host_galaxy_metallicity}

We have calculated the metallicity of the host galaxy of SN~2020rea using narrow emission line fluxes in the host galaxy spectrum taken on August 15, 2020 with LCO's FLOYDS spectrograph at Faulkes Telescope North (FTN). Prominent lines of H$\alpha$, [N {\sc II}], etc. are present in the host spectrum. There are several methods to measure the metallicity \citep{1991ApJ...380..140M, Kewley_2002, 10.1111/j.1365-2966.2004.07591.x, Pilyugin_2005}. These calculations involve flux measurements of various emission lines. Using the N2 index calibration of \cite{10.1111/j.1365-2966.2004.07591.x}, we estimate the  metallicity of the host galaxy as 12+log(O/H) = 8.56$\pm$0.18 dex. This is comparable to the metallicity of the host galaxy of SNe 2012Z (8.51$\pm$0.31 dex;  \citealt{2015ApJ...806..191Y}) and 2020sck (8.54$\pm$0.05 dex;  \citealt{2022ApJ...925..217D}). The metallicity measurements for host galaxy of faint Type Iax SNe such as SNe 2008ha, 2010ae, 2019gsc, 2020kyg are 8.16$\pm$0.15 dex \citep{2009AJ....138..376F}, 8.40$\pm$0.18 dex \citep{2014A&A...561A.146S}, 8.10$\pm$0.06 dex \citep{2020ApJ...892L..24S} and 8.68$\pm$0.04 dex \citep{2022MNRAS.511.2708S}, respectively. \cite{2017A&A...601A..62M}  demonstrated that there is no clear correlation between host galaxy metallicity and SN luminosity for Type Iax SNe, however with the increased sample we do see a tendency of Type Iax SNe to prefer metal poor hosts.

\section{Explosion scenario}
\label{explosion_sscenario}

SN~2020rea is one of the brightest members of Type Iax sub-class. In order to understand the most favorable explosion scenario for SN~2020rea, we compare the observational properties of SN~2020rea with different models one by one.

First, we consider the pulsational delayed detonation (PDD)
model. In the PDD scenario, the white dwarf remains bound while expanding due to slow deflagration and after that detonation occurs during pulsation because of compression and ignition caused by infalling C-O layers \citep{1974Ap&SS..31..497I,1991A&A...245L..25K,1991A&A...246..383K,1993A&A...270..223K,1995ApJ...444..831H,1996ApJ...457..500H,2006ApJ...642L.157B,Baron_2012,2014MNRAS.441..532D}. In the PDD explosion of a M$_{ch}$ C-O white dwarf, Fe group elements are produced in the deflagration phase. The mass of $^{56}$Ni produced in these model falls in between 0.12 to 0.66 M$_{\odot}$ \citep{1995ApJ...444..831H}. The estimated $^{56}$Ni mass for SN~2020rea matches with PDD5 model \citep{1995ApJ...444..831H} but ejecta velocity for SN~2020rea ($\sim$ 6500 km s$^{-1}$) is lower than that predicted by PDD5 model (8400 km s$^{-1}$). Also, the observed {\it (B-V)$_{0}$} colour at maximum ($-$0.01 mag) for SN~2020rea does not match with the {\it (B-V)$_{0}$} colour of PDD5 model (0.44 mag, \citealt{1995ApJ...444..831H}). 

Second, we consider a low energy core-collapse explosion model of a massive star which has been used to explain the observational features of some faint Type Iax SNe such as SN 2008ha \citep{2009Natur.459..674V,2009AJ....138..376F,2010ApJ...719.1445M}. Because of the low energy budget of faint SNe, a considerable amount of the ejecta falls back onto the remnant. This core-collapse scenario predicts kinetic energy of 1.2$\times$10$^{48}$ erg, 0.074 M$_{\odot}$ of ejecta mass and 0.003 M$_{\odot}$ of $^{56}$Ni \citep{2010ApJ...719.1445M} . Thus the predicted parameters in the core-collapse scenario are in disagreement with those of SN~2020rea. 

Next, we investigate the deflagration to detonation transition (DDT) model \citep{1991A&A...245L..25K,1991A&A...245..114K,1993A&A...270..223K,1995ApJ...444..831H,1996ApJ...457..500H,2002ApJ...568..791H,2013MNRAS.429.1156S,2013MNRAS.436..333S} which has been used to explain several observational properties of Type Ia SNe by varying the central density of white dwarf and strength of deflagration. The basic assumption in the deflagration to detonation models is that at late stage of explosion there is a transition of deflagration flame into a detonation front. DDT models \citep{2013MNRAS.429.1156S,2013MNRAS.436..333S} are generated by varying the number of ignition points.
The mass of $^{56}$Ni produced by these models (0.32 to 1.1 M$_{\odot}$, \citealt{2013MNRAS.436..333S}) is very high as compared to the $^{56}$Ni produced in SN~2020rea explosion. The range of kinetic energy (E$_{k}$ = 1.20-1.67 $\times$10$^{51}$ erg), absolute magnitude in {\it B}-band ($-$19.93 to $-$18.16 mag) and the redder {\it (B-V)$_{0}$} colour at maximum (0.15 to 0.56 mag) of the DDT models \citep{2013MNRAS.436..333S} do not agree with the estimated parameters of SN~2020rea.  

Finally, we take into account the three-dimensional pure deflagration of a C-O white dwarf \citep{2014MNRAS.438.1762F} which can successfully explain the observed properties of the bright and intermediate luminosity Type Iax SNe. These models provide a wide range of $^{56}$Ni mass between 0.03 to 0.38 M$_{\odot}$, rise time between 7.6 days to 14.4 days, and peak {\it V}-band absolute magnitudes spanning between $-$16.84 to $-$18.96 mag  \citep{2014MNRAS.438.1762F}. The observed parameters of SN~2020rea ($^{56}$Ni mass = 0.13$\pm$0.01 M$_{\odot}$, rise time = $\sim$ 16 days, {\it V}-band peak absolute magnitude = $-$18.30$\pm$0.12 mag) fall within the range prescribed by these models. In section \ref{bolometric_light_curve} we compared the pseudo-bolometric light curve of SN~2020rea with optical bolometric light curves presented in \cite{2014MNRAS.438.1762F}. The mixed abundance distribution given by these models is consistent with SN~2020rea. The expansion velocity inferred  from  Fe line is higher than Si lines indicating significant mixing in the ejecta. Furthermore, modelling the spectra of SN~2020rea with TARDIS (Section \ref{spectral_modelling}) suggests a mixed distribution of elements, consistent with the deflagration scenario.

\section{Summary}
\label{summary}

The photometric and spectroscopic investigations of SN~2020rea in optical wavelengths show that it lies at the brighter end of Type Iax luminosity distribution. The light curve decline rate in {\it B} and {\it g}-bands are $\Delta$m$_{15}$(B) = 1.61$\pm$0.14 mag and $\Delta$m$_{15}$(g) = 1.31$\pm$0.08 mag, respectively, indicating its similarity with SNe 2005hk and 2012Z. The colour evolution of SN~2020rea is analogous to other Type Iax SNe. Modeling of the pseudo bolometric light curve (constructed using {\it BgVri} bands) places SN~2020rea in the category of relatively bright Type Iax SNe with a rise time of  $\sim$ 16 days and $^{56}$Ni of 0.13$\pm$0.01 M$_{\odot}$. Assuming a photospheric velocity of 6500 km s$^{-1}$, ejecta mass and kinetic energy are estimated to be 0.77$^{+0.11}_{-0.21}$ M$_{\odot}$ and 0.19$^{+0.02}_{-0.06}$ $\times$ 10$^{51}$ erg, respectively. The comparison of the pseudo-bolometric light curve of SN~2020rea with optical bolometric light curves representing deflagration models of varying strength shows that the light curve of SN~2020rea is situated between N3-def and N5-def models during the early photospheric phase. The post-peak decline of the pseudo bolometric light curve is slower than the deflagration model light curves. The spectroscopic features of SN~2020rea are typical of Type Iax SNe. The Si {\sc II} line velocities of SN~2020rea are generally higher than those of other Type Iax SNe except for SNe 2011ay and 2012Z. The higher Fe line velocity than Si line around maximum indicates mixing of fully burned material. Spectral modelling  of SN~2020rea  shows weak Si {\sc II} feature in early photospheric phase, an IGEs dominated ejecta $\sim$ 10 days after maximum and hints towards a mixed ejecta. The host galaxy metallicity (8.56$\pm$0.18 dex) of SN~2020rea is similar to the host galaxy metallicity of SN 2012Z (8.51$\pm$0.31 dex). Out of the several proposed explosion scenarios for Type Iax SNe, pure deflagration of white dwarf emerges as a promising one to explain the observed properties of SN~2020rea.

\section*{Acknowledgments}
 We thank the anonymous referee for giving constructive comments which has improved the presentation of the paper. We acknowledge Wiezmann Interactive Supernova data REPository http://wiserep.weizmann.ac.il (WISeREP) \citep{2012PASP..124..668Y}. This research has made use of the CfA Supernova Archive, which is funded in part by the National Science Foundation through grant AST 0907903. This research has made use of the NASA/IPAC Extragalactic Database (NED) which is operated by the Jet Propulsion Laboratory, California Institute of Technology, under contract with the National Aeronautics and Space Administration. This work makes use of data obtained with the LCO Network. RD acknowledges funds by ANID grant FONDECYT Postdoctorado Nº 3220449. KM acknowledges BRICS grant DST/IMRCD/BRICS/Pilotcall/ProFCheap/2017(G) for the present work. The LCO group were supported by NSF Grants AST-1911151 and AST-1911225. This research made use of TARDIS, a community-developed software package for spectral synthesis in supernovae \citep{kerzendorf_wolfgang_2018_1292315, kerzendorf_wolfgang_2019_2590539}. The development of TARDIS received support from the Google Summer of Code initiative and from ESA's Summer of Code in Space program. TARDIS makes extensive use of Astropy and PyNE. This work made use of the Heidelberg Supernova Model Archive (HESMA)\footnote{\url{https://hesma.h-its.org}}.

\section*{ Data availability} The photometric and spectroscopic data of SN~2020rea presented in this paper will be made available by the corresponding author on request.
    











\bibliographystyle{mnras}
\bibliography{ms}

\bsp	
\label{lastpage}
\end{document}